\documentstyle[aps,aipbook,floats,psfig]{revtex}
\begin{document}
\title{The Status of Supersymmetry}

\author{Jonathan A. Bagger}
\address{Department of Physics \& Astronomy\\
Johns Hopkins University\\
Baltimore, MD\ 21218}
\maketitle

\begin{abstract}
\noindent
Supersymmetry searches are about to enter an important new era.  With
LEP 200 and the Main Injector, they will, for the first time, begin
to probe significant regions of the supersymmetric parameter space.
There is a real chance that supersymmetry might indeed be found
before the advent of the LHC.
\end{abstract}

\section*{Introduction}

\def\roughly#1{\raise.3ex\hbox{$#1$\kern-.75em\lower1ex\hbox{$\sim$}}}

Up to now, supersymmetry has been a theorist's dream and an
experimentalist's nightmare.  On the one hand, theorists tend
to like supersymmetry because it provides a beautiful
mathematical structure which can be used to stabilize
the mass hierarchy against radiative corrections.  On
the other hand, many experimentalists despise the subject
because supersymmetric predictions always seem to lie just
out of reach.

At present, direct searches for supersymmetry are just shots
in the dark because current accelerators do not have the power
to explore significant regions of the parameter
space.  As we will see, this will soon change, but for now,
direct searches do not restrict the theory in any
important way.

Precision electroweak measurements also reveal very little
about supersymmetry.  The technical reason for this is
that supersymmetry decouples from all standard-model electroweak
observables.  For example, the supersymmetric and standard-model
values of $S$ and $T$ are related as follows \cite{Peskin}
\begin{eqnarray}
S_{\rm SUSY} &\ =\ & S_{\rm SM}\ +\ {\cal O}(M_W/M_S)^2 \nonumber \\
T_{\rm SUSY} &\ =\ & T_{\rm SM}\ +\ {\cal O}(M_W/M_S)^2 \nonumber\ ,
\end{eqnarray}
where $M_W$ is the mass of the $W$, and $M_S$ denotes the scale
of the supersymmetric spectrum.  Theorists can simply raise
$M_S$ and bring supersymmetry into complete accord with
standard-model predictions.

Fortunately, the next generation of accelerators, including the
Fermilab Main Injector, LEP 200, and a possible higher-luminosity
Tevatron, will open a new era in supersymmetric particle searches.
These accelerators will -- for the first time -- begin to probe
significant regions of the supersymmetric parameter space.  And
with the advent of the LHC, the search for supersymmetry will
finally cover most -- if not all -- of the parameter space that
is relevant for weak-scale supersymmetry.

In this talk we will examine the supersymmetric parameter space.
We will impose supersymmetric unification and show the preferred range
for the supersymmetric particle masses.  We will also discuss two more
specialized points:  the constraints on the supersymmetric spectrum
that follow from supersymmetric unification, as well as the one-loop
constraints on the mass of the lightest Higgs boson.

\section*{The Supersymmetric Spectrum}

Supersymmetry stabilizes the hierarchy $M_W/M_P \simeq 10^{-17}$
against quadratically-divergent radiative corrections.  However, it
does so at a tremendous cost:  a doubling of the particle spectrum.
In supersymmetric theories, all bosons are paired with fermions
and vice versa.  For the case at hand, each of the standard-model
particles is accompanied by a supersymmetric partner with the same
SU(3) $\times$ SU(2) $\times$ U(1) quantum
numbers.  If supersymmetry is not broken, these superparticles are
degenerate in mass with their original standard-model partners.

\begin{figure}
\vbox{
\centerline{\psfig{figure=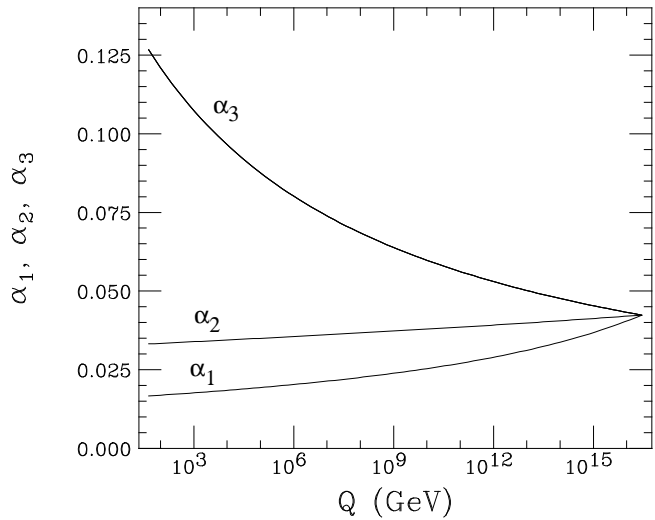,height=3.0in}}
\caption{The gauge couplings unify in the minimal supersymmetric
standard model.}
}
\label{figlabel}
\end{figure}

The fact that such particles have not been observed tells us that
supersymmetry must be broken -- and that the breaking must be soft.
In other words, the supersymmetry breaking must not reintroduce
destabilizing quadratic divergences.  The various types of soft
supersymmetry breaking have been studied extensively \cite{Girardello}.
One finds over 50 new parameters:

\begin{itemize}
\item 3 gaugino masses:  $M^{(a)}_{1/2}$
\item 23 scalar masses: $M^2_{0\ ij^*}$
\item 27 trilinear scalar couplings:  $A_{0\ ijk}$
\item 1 bilinear scalar coupling:  $B \mu$.
\end{itemize}

\noindent
(plus phases).  Naturalness requires that each of the dimensionful
couplings be less than about a TeV, but that is all we know.  The most
general softly-broken supersymmetric theory has an enormous parameter
space -- and theorists are prepared to use every bit of it to evade
experimental limits!

\begin{figure}
\vbox{
\centerline{\psfig{figure=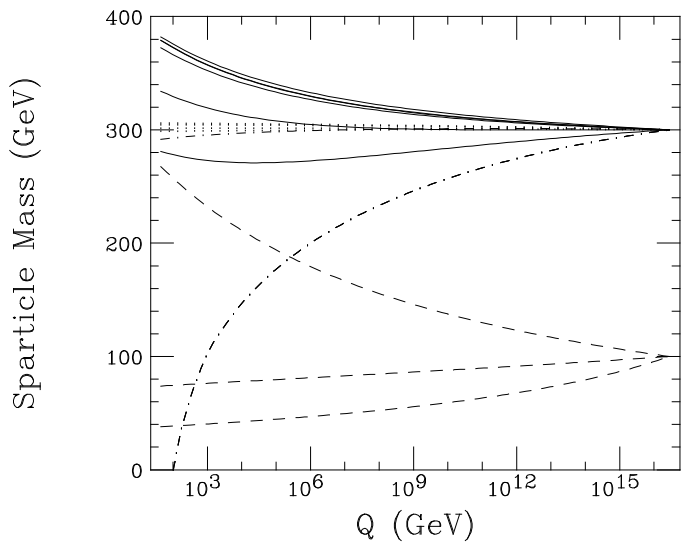,height=3.0in}}
\caption{The soft supersymmetry-breaking masses can be arranged to
unify in the supersymmetric standard model.  Here $M_0 = 300$ GeV
and $M_{1/2} = 100$ GeV.  The solid lines denote squark masses and
the dotted lines sleptons. The dashed lines are gaugino masses,
while the dot-dashed line marks the mass of the Higgs.}
}
\label{figlabel}
\end{figure}

Motivated by the successful unification of the gauge couplings in the
supersymmetric standard model (Fig.~1), it is not unreasonable to assume
that the soft breakings unify as well.  In this case the parameter space
reduces substantially, to include

\begin{itemize}
\item 1 universal gaugino mass:  $M_{1/2}$
\item 1 universal scalar mass: $M^2_0$
\item 1 universal trilinear scalar coupling:  $A_0$
\item 1 bilinear scalar coupling:  $B \mu$.
\end{itemize}

\noindent
As usual in a unified theory, these parameters are fixed at the
unification scale $M_{GUT}$.  Their values in the low-energy
theory are determined by the renormalization group, as shown
in Fig.~2.

At the unification scale, the unification assumption forbids
electroweak symmetry breaking because all scalar masses -- including
that of the Higgs -- have the common value $M^2_0$.  However,
top-quark loops decrease the Higgs mass.  Therefore, in the
renormalization group evolution, they drive down the mass of
the Higgs.  If the top Yukawa coupling is sufficiently
large (corresponding to a top mass of about 150 -- 200 GeV),
the mass squared goes negative and triggers electroweak symmetry
breaking.  It is remarkable that this radiative mechanism for
symmetry breaking \cite{Ibanez}, first proposed in 1983, is
based on a top-quark mass that is in complete agreement with
current experiments!

In what follows, we present expectations for the supersymmetric
spectrum based on this unification scenario.  For simplicity, we
set $A_0 = 0$, and we trade $B$ for the value of $\tan\beta =
v_d/v_u$, evaluated at the scale $M_Z$.  In our figures we take
$\alpha_s(M_Z) = 0.12$, $M_t = 175$ GeV, and the supersymmetric
Higgs mass parameter $\mu > 0$.  We find the values of various
supersymmetric masses as a function of $M_0$ and $M_{1/2}$, for
two values of $\tan\beta$.  (In the figures, all masses are
one-loop pole masses.)

\begin{figure}
\vbox{
\centerline{\psfig{figure=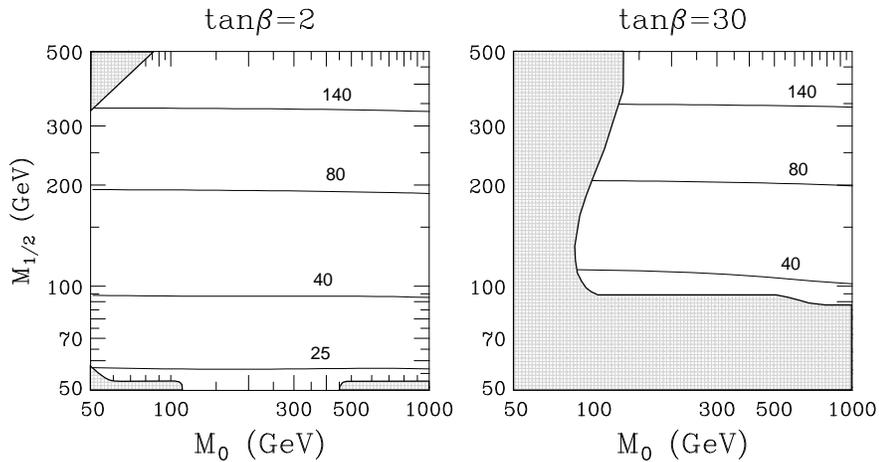,height=2.75in}}
\caption{The mass of the lightest supersymmetric particle,
$\chi^0_1$, for $\mu > 0$, $A_0 = 0$, $\alpha_s(M_Z) =
0.12$ and $M_t = 175$ GeV.  The shaded region is forbidden
by experimental and theoretical constraints.  Most of the
supersymmetric parameter space is still open.}
}
\label{figlabel}
\end{figure}

In Fig.~3 we show mass contours for the lightest superparticle,
$\chi^0_1$.  The $\chi^0_1$ is neutral, and because of a global
symmetry called $R$-parity, is assumed to be stable.  In the figure,
the shaded areas represent forbidden regions of parameter space, either
because of present experimental limits or because of theoretical
constraints such as the cosmological requirement that the lightest
(stable) superparticle be neutral, or the phenomenological constraint
that electroweak symmetry be broken, but not color.

\begin{figure}
\vbox{
\centerline{\psfig{figure=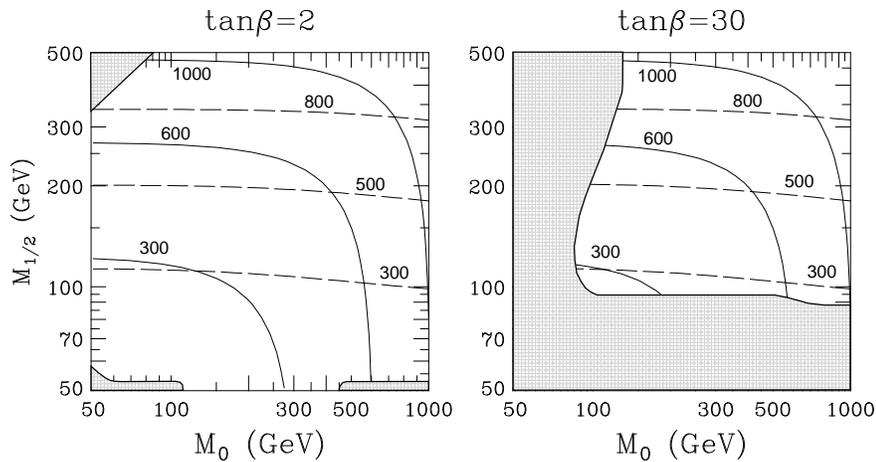,height=2.75in}}
\caption{The mass of the up squark (solid line) and the
gluino (dashed line), for $\mu > 0$, $A_0 = 0$, $\alpha_s(M_Z)
= 0.12$ and $M_t = 175$ GeV.  Our parameter space
corresponds to $M_{\tilde q}\ \roughly{<}\ 1$ TeV.}
}
\label{figlabel}
\end{figure}

In Fig.~4 we show contours for the (up) squark and the gluino
masses.  (The masses of the up, down, charm and strange squarks
are almost degenerate.)  From the plot we see that our parameter
space covers squark masses up to about 1 TeV.  This is the range of
interest if supersymmetry is to stabilize the weak-scale hierarchy.
(The rule of thumb is that $M_{\tilde g} \sim 3 M_{1/2}$
and  $M^2_{\tilde q} \sim M^2_0 + 4 M^2_{1/2}$.)

\begin{figure}
\vbox{
\centerline{\psfig{figure=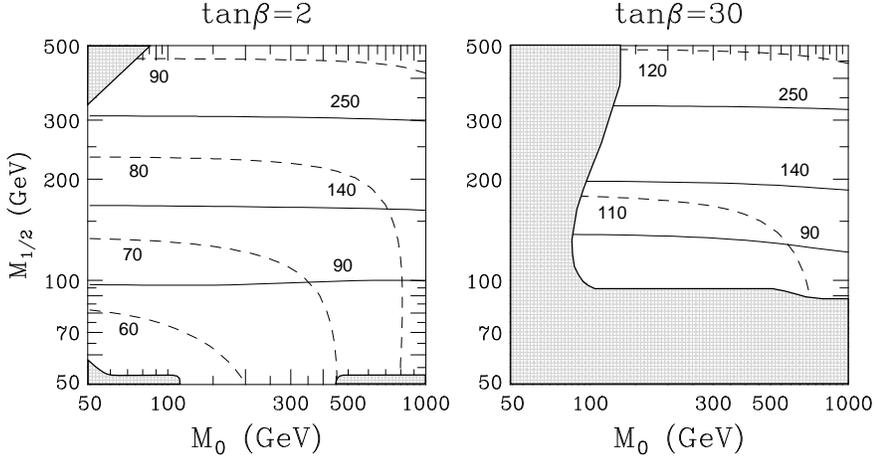,height=2.75in}}
\caption{The mass of the lightest chargino, $\chi^\pm_1$,
(solid line) and lightest Higgs, $h$, (dashed line), for $\mu
> 0$, $A_0 = 0$, $\alpha_s(M_Z) = 0.12$ and $M_t = 175$ GeV.
The Higgs mass $M_h\ \roughly{<}\ 120$ GeV over our
parameter space.}
}
\label{figlabel}
\end{figure}

In Fig.~5 we plot contours for the masses of the lightest Higgs
scalar, $h$, and the lightest chargino, $\chi^\pm_1$.  We see
that $M_{\chi^\pm} \sim  M_{1/2}$, and that for our parameter
space, the maximum Higgs mass is about 120 GeV.  (For completeness,
we note that the slepton masses are approximately $M_{\tilde \ell}
\sim M_0$.)

\begin{figure}
\vbox{
\centerline{\psfig{figure=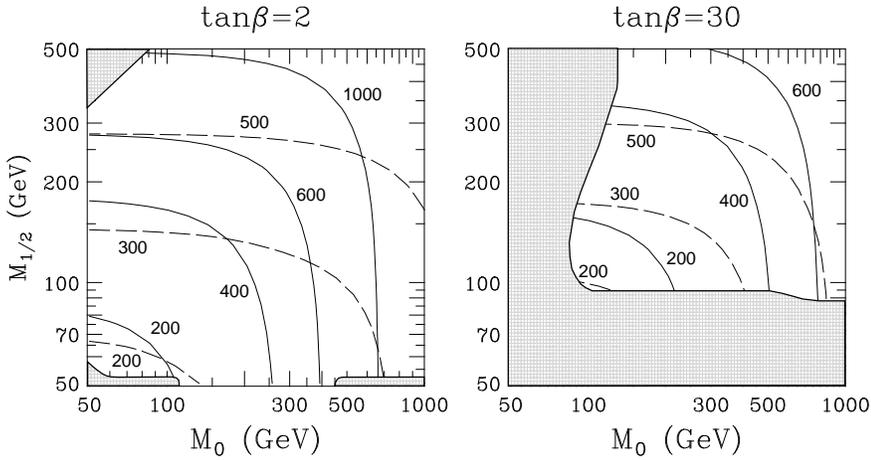,height=2.75in}}
\caption{The mass of the charged Higgs, $H^\pm$,
(solid line) and lightest stop, $\tilde t$, (dashed line),
for $\mu > 0$, $A_0 = 0$, $\alpha_s(M_Z) = 0.12$ and $M_t =
175$ GeV.  The decays $t \rightarrow
\tilde t \tilde \chi^0_1$ and $t \rightarrow H^+ b$ are
kinematically forbidden over most of the parameter space.}
}
\label{figlabel}
\end{figure}

Finally, in Fig.~6 we show contours for the
lightest stop squark, $\tilde t$, and charged Higgs,
$H^\pm$.  From the figure we see that the decays $t \rightarrow
\tilde t \tilde \chi^0_1$ and $t \rightarrow H^+ b$ are kinematically
forbidden over most of the parameter space.  (The stop can be
lighter for $A_0 \ne 0$, but a very light stop requires a fine
tuning of the parameters.)

These figures can be used to illustrate the supersymmetry reach
of a given accelerator.  For example, LEP 200 has a mass reach
of about $\sqrt s - 100$ GeV for a supersymmetric Higgs particle,
and $\sqrt s /2$ for a chargino.  Therefore Fig.~5 shows that
LEP 200 has an excellent chance of discovering the lightest
supersymmetric Higgs and a reasonable possibility of finding the
lightest chargino.

\begin{figure}
\vbox{
\centerline{\psfig{figure=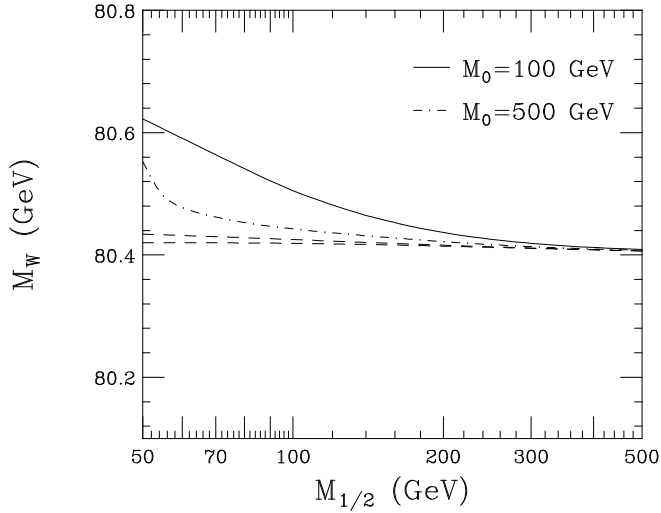,height=3.0in}}
\caption{The one-loop $W$-boson pole mass, for $\tan\beta = 2$,
$A_0 = 0$, $\alpha_s(M_Z) = 0.12$, $M_t = 175$ GeV, and two values
of $M_0$.  The dashed lines correspond to the standard-model
$W$ masses, assuming the same Higgs masses as in the
supersymmetric cases.  Note that the supersymmetry effects
decouple for large $M_{1/2}$.}
}
\label{figlabel}
\end{figure}

The Tevatron's discovery potential is more model-dependent, and
varies considerably with the Tevatron luminosity.  For an integrated
luminosity between 200 pb${}^{-1}$ and 25 fb${}^{-1}$, the gluino
discovery reach is in the range of 300 -- 400 GeV.  Likewise, the
chargino/neutralino reach varies between 150 -- 250 GeV in the
trilepton decay channel, $\chi^+_1 \chi^0_2 \rightarrow \ell^+
\ell^- \ell^{\prime +}$ plus missing energy \cite{Baer}.  From
Figs.~4 and 5 we see that an upgraded Tevatron would begin
to cover a significant amount of the supersymmetric parameter
space.

\section*{Radiative Corrections}

In the introduction, we argued that precision electroweak measurements
cannot be used to restrict the allowed regions of $M_0$ and $M_{1/2}$
because supersymmetric effects decouple from electroweak
observables.  This can be seen explicitly in Fig.~7, where we plot
the one-loop $W$-boson pole mass against $M_{1/2}$, for two values
of $M_0$.  We see that the supersymmetric $W$ mass is indistinguishable
from the standard-model mass for $M_{1/2}\ \roughly{>}\ 300$ GeV.  The
figure also shows that a measurement uncertainty for $M_W$ of 40 MeV
gives a sensitivity to the region $M_{1/2}\ \roughly{<}\ 150$ GeV for
$M_0 \simeq 100$ GeV.  But this is just the region that will be
probed directly by LEP 200 and by the Tevatron with the Main
Injector!

In the context of supersymmetric unification, however, precision
measurements can play an important role in restricting the
supersymmetric parameter
space.  To see this, we recall that in a unified model, $\alpha_s
(M_Z)$ can be predicted as a function of $\alpha_1(M_Z)$, $\alpha_2
(M_Z)$, and the weak- and unification-scale thresholds.  For the
case at hand, we determine $\alpha_1(M_Z)$ and $\alpha_2
(M_Z)$ from the electroweak observables $\alpha_{EM}$, $G_F$ and
$M_Z$.  We then calculate the weak-scale thresholds using the
minimal supersymmetric standard model, and take the
unification-scale thresholds to be those of a particular unified
model.  In this way we can compute $\alpha_s(M_Z)$ as a function
of the parameters $M_0$ and $M_{1/2}$ in any unified model.

\begin{figure}
\vbox{
\centerline{\psfig{figure=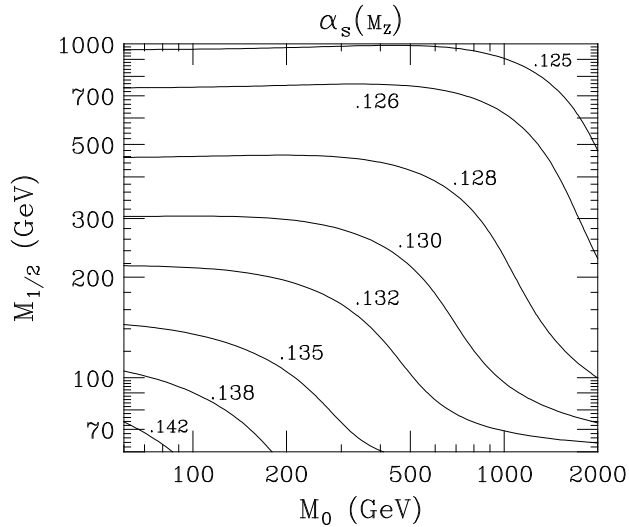,height=3.0in}}
\caption{Contours of $\alpha_s(M_Z)$ in the $M_0$, $M_{1/2}$
plane with $m_t=175$ GeV, $\tan\beta=2$ and $A_0=0$.}
}
\label{figlabel}
\end{figure}

In Fig.~8 we show the prediction for $\alpha_s(M_Z)$ in the
absence of unification-scale thresholds.  We see that $\alpha_s(M_Z)$
is generally much larger than the experimental value of $\alpha_s(M_Z)
= 0.117 \pm 0.005$.  Indeed, we see that $\alpha_s(M_Z) \le 0.127$
requires $M_0\ \roughly{>}\ 1$ TeV or $M_{1/2}\ \roughly{>}\ 500$
GeV \cite{Bagger}.

\begin{figure}
\vbox{
\centerline{\psfig{figure=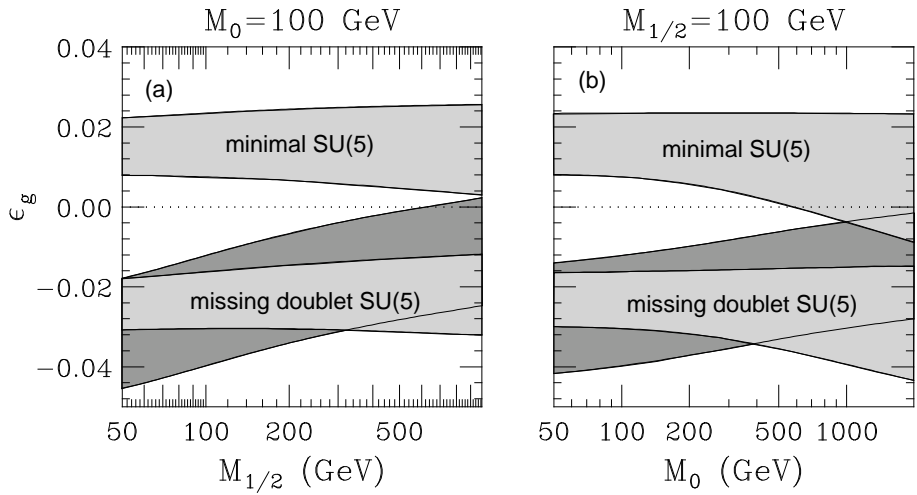,height=3.0in}}
\caption{The light shaded regions indicate the allowed
values of the gauge coupling threshold correction
$\epsilon_g$ in the minimal and missing-doublet SU(5)
models.  The dark shaded region indicates the range of
$\epsilon_g$ necessary to obtain $\alpha_s(M_Z) =
0.117 \pm 0.01$.}
}
\label{figlabel}
\end{figure}

In Fig.~9 we show the effects of unification-scale thresholds.
We parametrize these thresholds by $\epsilon_g$, and illustrate
the allowed values in the minimal and missing-doublet SU(5) models,
together with the threshold required to give $\alpha_s(M_Z) = 0.117
\pm 0.01$.  From the figure we see that minimal SU(5) requires
$M_0\ \roughly{>}\ 1$ TeV, which leads to squark masses of more than 1
TeV.

Weak-scale radiative corrections are also important
in determining the Higgs mass.  As is well-known,
in supersymmetric models the tree-level Higgs mass is determined by
gauge couplings, and is bounded from above by $M_Z$.  For heavy
top, this value receives significant radiative corrections, and for
$M_t \simeq 175$ GeV, the bound increases to about 120 GeV.

Experimentally, this is a very interesting number because it is
almost within reach of LEP 200.  Theoretically, $M_h \simeq 120$
GeV is interesting as well, because it is approximately the {\it lower}
bound for the Higgs mass in the ordinary, nonsupersymmetric standard
model.  In the standard model, top-quark loops give a negative
logarithmically-divergent $\phi^4$ contribution to the effective
potential, and this contribution can destabilize the vacuum.  If
we require that the standard model hold all the way
to the Planck scale, so that the cutoff $\Lambda \simeq M_P$,
then the Higgs mass must be more than about 120 GeV \cite{Quiros}.

\begin{figure}
\vbox{
\centerline{\psfig{figure=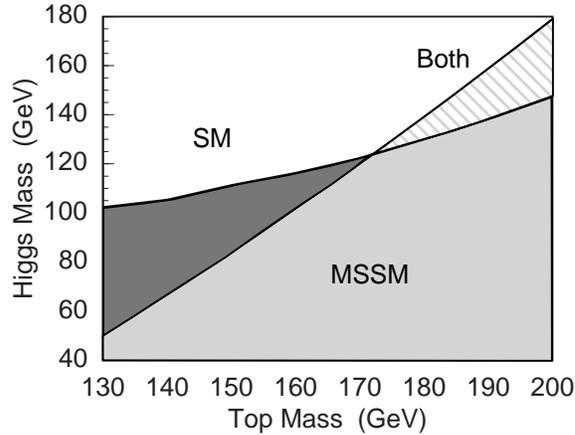,height=2.8in}}
\caption{The maximum one-loop Higgs mass in the minimal
supersymmetric standard model, and the minimum Higgs mass
in the ordinary standard model, as a function of the
top-quark mass (After Ref.~(7)).}
}
\label{figlabel}
\end{figure}

This is illustrated in Fig.~10, where we plot the allowed
Higgs masses as a function of $M_t$.  From the figure we see
that the maximum mass increases with $M_t$ in supersymmetric
standard model, as does the minimum mass in the ordinary
standard model.  The curves have different slopes, and cross
at $M_t \simeq 175$ GeV.  These curves indicate that
if the Higgs is discovered at LEP 200, either supersymmetry
is right, or that there must be other new physics below the
Planck scale!

\section*{Conclusions}

In this talk we have seen that supersymmetry searches are about to
enter an important new era.  With LEP 200 and the Main Injector, they
will begin to probe large regions of the supersymmetric parameter
space.  With luck, supersymmetry might even be found before the
advent of the LHC!

I would like to thank my collaborators Konstantin Matchev, Renjie
Zhang, and especially Damien Pierce for sharing their insights
on the supersymmetric standard model.  This work was supported
by the U.S. National Science Foundation under grant NSF-PHY-9404057.

\end{document}